\documentclass[conference]{IEEEtran}
\usepackage{blindtext}

\usepackage[T1]{fontenc}
\usepackage[utf8]{inputenc}
\usepackage[english]{babel}

\usepackage{float} 

\usepackage{xcolor} 
\definecolor{lightgray}{gray}{0.6}
\definecolor{medgray}{gray}{0.4}

\usepackage{hyperref}
\hypersetup{
	colorlinks=true,
	urlcolor= black,
	citecolor=black,
	linkcolor= black,
	bookmarks=true,
	bookmarksopen=false,
}

\newif\ifptitle
\newif\ifpnumber
\newcounter{para}
\newcommand\ptitle[1]{\par\refstepcounter{para}
	{\ifpnumber{\noindent\textcolor{lightgray}{\textbf{\thepara}}\indent}\fi}
	{\ifptitle{\textbf{[{#1}]}}\fi}}
\ptitletrue  
\pnumbertrue  

\usepackage{graphicx}
\graphicspath{ {images/} }

\usepackage{ragged2e} 
\usepackage{csquotes} 

\usepackage{listings}
\usepackage{xcolor,colortbl}
\definecolor{myblue}{RGB}{0,163,243}
\definecolor{mygrey}{RGB}{128,128,128}
\definecolor{whitesmoke}{RGB}{245,245,245}
\usepackage{tcolorbox} 
\newtcolorbox[auto counter, number within=section]{mybox}[2][]{
	colbacktitle=mygrey,
	colback=whitesmoke,
	colframe=mygrey,  
	coltitle=white,   
	title={#2},       
	fonttitle=\ttfamily\small,
	fontupper=\sffamily\small,
	halign=flush left,
	rounded corners
}

\begin{document}
\title{Optimizing Spot Instance Reliability and Security Using Cloud-Native Data and Tools}

\author{
	\IEEEauthorblockN{Muhammad Saqib}
	\IEEEauthorblockA{
		Texas Tech University\\
		Dept. of Computer Science\\
		\texttt{saqibraopk@hotmail.com}
	}
\and
\IEEEauthorblockN{Shubham Malhotra}
\IEEEauthorblockA{
    Rochester Institute of Technology\\
    Dept. of Software Engineering\\
    \texttt{shubham.malhotra28@gmail.com}
}
\and
\IEEEauthorblockN{Dipkumar Mehta}
\IEEEauthorblockA{
    C.K.Pithawalla College of Eng. and Tech.\\
    \texttt{dipkumar.mehta@gmail.com}
}
\and
\IEEEauthorblockN{Jagdish Jangid}
\IEEEauthorblockA{
    Infinera Corp\\
    \texttt{jangid.jagdish@gmail.com}
}
\and
\IEEEauthorblockN{Fnu Yashu}
\IEEEauthorblockA{
    Stony Brook University\\
    Dept. of Computer Science\\
    \texttt{yyashu@cs.stonybrook.edu}
}
\and
\IEEEauthorblockN{Sachin Dixit}
\IEEEauthorblockA{
    Stripe Inc.\\
    \texttt{spdixit@gmail.com}
}
}

\maketitle

\begin{abstract}
  This paper presents "Cloudlab," a comprehensive, cloud-native laboratory designed to support network security research and training. Built on Google Cloud and adhering to GitOps methodologies, Cloudlab facilitates the creation, testing, and deployment of secure, containerized workloads using Kubernetes and serverless architectures. The lab integrates tools like Palo Alto Networks firewalls, Bridgecrew for "Security as Code," and automated GitHub workflows to establish a robust Continuous Integration/Continuous Machine Learning pipeline. By providing an adaptive and scalable environment, Cloudlab supports advanced security concepts such as role-based access control, Policy as Code, and container security. This initiative enables data scientists and engineers to explore cutting-edge practices in a dynamic cloud-native ecosystem, fostering innovation and improving operational resilience in modern IT infrastructures.  
\end{abstract}

\section{\label{sec:Start}Introduction}
\vspace{2mm}
\justifying

The ever-existing nature of cloud-native technologies has brought a shift in the way organizations build, deploy, and manage modern applications. As businesses switch to cloud environments—spanning public, private, and hybrid models—they gain access to advanced scalability and flexibility. However, this transition introduces a host of complexities, particularly in ensuring the reliability, security, and adaptability of cloud-based solutions. Organizations must steer their way through an ever-expanding ecosystem of tools, frameworks, and methodologies to address challenges such as securing containerized workloads, managing role-based access, implementing continuous integration, and adhering to Policy as Code principles. These challenges are further compounded by the dynamic nature of cloud-native environments, where traditional security paradigms often fall short.

To address these pressing concerns, this paper introduces Cloudlab, a dedicated cloud-native laboratory meticulously designed to support advanced security research, testing, and training. Built on Google Cloud's infrastructure and adhering to GitOps methodologies, Cloudlab serves as a comprehensive platform for exploring and validating cutting-edge security concepts. It incorporates advanced technologies such as Kubernetes, serverless architectures, and automated workflows to create a scalable and adaptive environment. The lab integrates tools like Palo Alto Networks CN-Series firewalls and Bridgecrew's "Security as Code" framework, enabling researchers and engineers to adopt best practices in cloud security.

The foundation of Cloudlab lies in its dual pipelines: Continuous Integration (CI) and Continuous Machine Learning (CML). The CI pipeline is responsible for generating Docker images, managing version control, and automating the testing and deployment of secure workloads. On the other hand, the CML pipeline facilitates experimentation with machine learning models and their secure deployment. Together, these pipelines foster an ecosystem of automation, innovation, and operational resilience, aligning with the needs of modern IT infrastructures.

Through Cloudlab, this paper seeks to explore a range of topics critical to cloud-native environments. These include provisioning secure infrastructure, implementing serverless cloud functions, automating security workflows, and ensuring the security of containerized applications. The lab's design not only promotes the adoption of advanced security practices but also bridges the gap between theoretical research and real-world application.

By delving into the nuances of API endpoints, container security, role-based access control, and Policy as Code, Cloudlab provides a hands-on platform for engineers and security practitioners to enhance their expertise. Furthermore, the lab underscores the importance of automating security processes, leveraging tools like Terratest and Kyverno to validate and enforce security policies. These capabilities demonstrate the potential of cloud-native approaches to revolutionize the field of network security, offering robust solutions to the challenges posed by modern cloud ecosystems.

This paper aims to provide a detailed account of Cloudlab's architecture, capabilities, and applications, showcasing how it addresses key challenges in cloud security and reliability. By presenting a scalable, adaptable, and secure laboratory environment, this study highlights the transformative potential of cloud-native tools and methodologies. Ultimately, Cloudlab serves as a testament to the critical role of innovation in advancing security practices and supporting the next generation of cloud-native technologies.\cite{CloudShift}.

\justifying
The Cloudlab is a cutting-edge, private laboratory environment meticulously designed to advance security research, testing, and training in cloud-native ecosystems. It represents a paradigm shift in how modern security practices are approached, leveraging the principles of cloud-native technologies to ensure scalability, adaptability, and operational efficiency. What sets Cloudlab apart is its adherence to GitOps practices \href{https://www.cloudbees.com/gitops/what-is-gitops}, a revolutionary methodology that redefines infrastructure management. By treating infrastructure as code (IaC), GitOps facilitates the seamless storage, review, and maintenance of complex cloud configurations, ensuring a robust foundation for experimentation and deployment.

At its core, Cloudlab is engineered to support researchers, engineers, and practitioners in tackling the multifaceted challenges of securing modern IT infrastructures. It does so through two primary, purpose-driven pipelines. The first pipeline is dedicated to Continuous Integration (CI), streamlining the development process by automating the testing, building, and deployment of containerized workloads. This pipeline ensures that new code integrations are rigorously tested and securely deployed, fostering a culture of rapid innovation without compromising stability.

The second pipeline, \href{https://cml.dev/doc}{Continuous Machine Learning}, or CML\cite{books/mit/026233758}, is equally transformative. As artificial intelligence (AI) and machine learning (ML) become integral to cloud-native systems, the CML pipeline facilitates the development, testing, and secure deployment of ML models. By automating the end-to-end lifecycle of machine learning workflows, the CML pipeline enables researchers to iterate rapidly, test at scale, and deploy models with a high degree of reliability. This capability is particularly valuable for exploring novel AI-driven security solutions, enhancing the lab’s capacity for cutting-edge research.

Together, these pipelines form the backbone of Cloudlab, creating a dynamic and adaptive environment capable of addressing the demands of modern security challenges. They empower users to experiment with advanced concepts like role-based access control (RBAC), Policy as Code, and container security while integrating state-of-the-art tools and methodologies. Cloudlab’s adherence to GitOps practices ensures that every change is version-controlled and reproducible, providing a transparent and collaborative platform for innovation.

This holistic approach makes Cloudlab an invaluable resource for security practitioners aiming to stay ahead in the rapidly evolving world of cloud-native technologies. By combining rigorous methodologies with advanced tools, the lab not only supports groundbreaking research but also bridges the gap between theoretical knowledge and practical application. In doing so, it paves the way for a more secure and resilient future in cloud computing. 

\justifying
As security practitioners, it is important for us to understand the infrastructure and applications being built in the public
clouds as a first step to making things more secure.

\section{\label{sec:Project}Project Goals}
\vspace{2mm}
\justifying
There are several goals associated with the development and operation of this project, each of which plays a pivotal role in achieving a comprehensive understanding of cloud-native environments. These goals are interconnected, emphasizing the integration of advanced tools, methodologies, and security concepts. The overarching objective is to explore, evaluate, and enhance the composition, management, and resilience of various components, as outlined below:

\begin{raggedright}
	\begin{enumerate}
		\item Provisioning \href{https://docs.paloaltonetworks.com/cn-series.html}{Palo Alto Networks CN-Series firewall} products, integrating them with \href{https://www.tigera.io/project-calico/}{Calico} and protecting containerized workloads (Kubernetes ``pods'').
		\item Researching containerized deployments, workloads, and security.
		\item Integration of \href{https://docs.bridgecrew.io/docs}{Bridgecrew ``Security as Code'' tooling} with GitHub repositories.
		\item Developing and demonstrating expertise in ``serverless cloud functions'' (GCP Cloud Functions in this case).
		\item Developing and utilizing a cloud-native Continuous Integration build pipeline. The output of this pipeline is a Docker image that is stored in gcr.io. These images include a fully contained set of tools, documentation, and Terraform code for customer deployments.
		\item Demonstrating Policy as Code concepts using \href{https://www.accurics.com/products/terrascan/}{Terratest} and \href{https://kyverno.io/}{Kyverno}.
	\end{enumerate}
\end{raggedright}

\vspace{2mm}

\section{\label{sec:LO}Learning Objectives}
\vspace{2mm}

\justifying
The primary aim of building and operating this lab is to foster learning and experimentation in key areas of cloud-native security. Engineers and researchers can benefit from engaging with various topics that are critical to modern cloud operations. The following list highlights the main learning objectives:

\begin{raggedright}
	\begin{enumerate}
		\item API Endpoints.
		\item Deployment and operation of CN-Series firewalls.
		\item Kubernetes role-based access control (RBAC) and security.
		\item Developing serverless functions in Python.
		\item ML/AI pipelines, containerized workloads, and their security.
		\item Containers and container security.
		\item Automation.
		\item Testing perspectives, including Policy as Code and Security as Code.
		\item CI/CD Pipelines and security.
	\end{enumerate}
\end{raggedright}

\vspace{2mm}

\justifying
Training materials derived from the lab’s construction and operation are readily available, enabling engineers to build proficiency in these areas and apply their knowledge to real-world scenarios.

\subsection{\label{sec:apis}API Endpoints}
\vspace{2mm}

\justifying
API endpoints are foundational to the interaction of applications across environments, including serverless systems and Kubernetes clusters. Their construction and operation provide an ideal space for exploring potential vulnerabilities and security challenges within cloud-native ecosystems. Understanding and securing API endpoints is critical for maintaining the reliability and integrity of applications, as they often serve as gateways for data exchange and system integration.

\subsection{\label{sec:contsec}Containers and Container Security}
\vspace{2mm}

\justifying
Containerized environments, while offering immense flexibility and scalability, introduce unique security challenges. Addressing these challenges involves understanding container architecture, identifying vulnerabilities, and implementing best practices for container security. This includes securing Kubernetes pods, employing role-based access controls, and integrating automated tools to monitor and protect containerized workloads.

\subsection{\label{sec:auto}Automation}
\vspace{2mm}

\justifying
Automation is a cornerstone of cloud-native operations, enabling organizations to scale efficiently and maintain operational consistency. By automating processes such as infrastructure provisioning, security scans, and policy enforcement, organizations can significantly reduce manual errors and improve overall system reliability. The lab emphasizes the use of tools and methodologies to automate complex workflows, ensuring that security and operational excellence remain integral to cloud-native environments.

\section{\label{sec:Lab}Lab Configuration Details}
\vspace{2mm}

\justifying
\justifying
The configuration and codebase for the lab are meticulously maintained on GitHub, ensuring transparency, collaboration, and version control. By leveraging GitHub repositories as a centralized storage system, the lab facilitates efficient infrastructure management, version tracking, and seamless collaboration between engineers, researchers, and developers. Hosted on Google Cloud—one of the ``big three'' public cloud providers—the lab serves a dual purpose: acting as a proof of concept for innovative cloud-native solutions and as a training platform to enhance hands-on expertise in modern security and automation practices.

The lab’s architecture is composed of several core components that work cohesively to create a secure, scalable, and adaptive environment. Each of these components is described below:

\begin{enumerate}
	\item \textbf{GitHub Repositories}: The GitHub repositories are the foundation of the lab's infrastructure as code (IaC) approach. They store all configuration files, pipeline scripts, and containerized application code, ensuring a single source of truth. By maintaining these files in GitHub, the lab enables automated workflows such as pull request reviews, code scanning, and testing, which are crucial for ensuring the reliability and integrity of the infrastructure. These repositories also facilitate collaboration by allowing multiple contributors to work simultaneously, with built-in features for managing changes and resolving conflicts.

	\item \textbf{Google Kubernetes Engine (GKE)}: GKE provides the backbone for the lab’s computational and orchestration needs. It hosts the Kubernetes clusters that underpin the Continuous Integration (CI) pipeline and enable machine learning experimentation. The GKE clusters are designed to support containerized workloads efficiently, ensuring seamless scalability and resource optimization. By integrating role-based access control (RBAC) and network policies, the GKE environment ensures that workloads are both secure and compliant with best practices. This infrastructure allows for real-time testing and deployment of workloads while maintaining high availability and fault tolerance.

	\item \textbf{Private Container Repository (GCR)}: The Google Container Registry (GCR) serves as a secure storage location for Docker images generated by the CI pipeline. These images encapsulate the tools, libraries, and application code required for deployment, ensuring consistency across environments. By maintaining a private repository, the lab guarantees that sensitive configurations and dependencies are securely stored, reducing the risk of unauthorized access. Additionally, GCR supports automated vulnerability scanning, allowing the lab to identify and address potential security risks in container images.

	\item \textbf{Automation Frameworks}: Automation is a cornerstone of the lab’s design, with frameworks such as GitHub Actions and webhooks playing a critical role in streamlining CI/CD workflows. These automation tools trigger events such as testing, building, and deployment whenever changes are made to the codebase. Webhooks enable seamless integration between GitHub repositories and other services, such as the CI pipeline hosted on GKE. The automation frameworks also incorporate security measures like ``Security as Code'' tooling, ensuring that every build and deployment adheres to predefined security standards.

\end{enumerate}

\justifying
The lab’s design underscores the importance of integrating automation, security, and scalability into cloud-native environments. By harmonizing these elements, the lab creates a robust platform for exploring advanced cloud concepts and gaining practical expertise. It empowers researchers and engineers to test innovative solutions in a controlled environment, bridging the gap between theoretical knowledge and real-world application.

\begin{figure}[ht]
	\centering
	\includegraphics[width=0.8\linewidth]{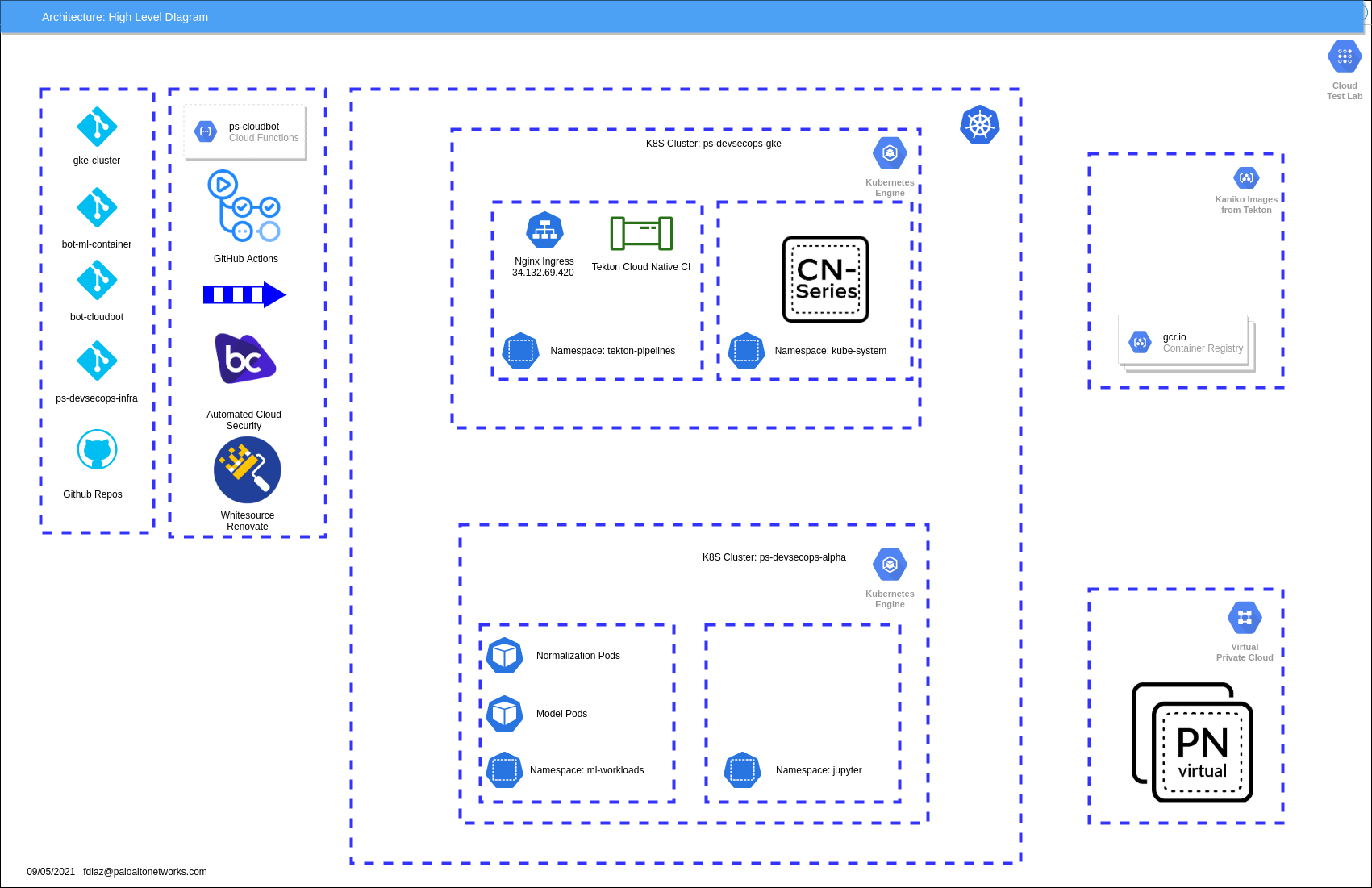}
	\caption{High-Level Lab Design Diagram}
	\label{fig:design}
\end{figure}

\justifying
A detailed description of the five main blocks seen in Figure~\ref{fig:design} is provided below:

\begin{enumerate}
	\item \textbf{CI/CD Pipelines}: The lab incorporates a robust CI/CD pipeline hosted on the GKE cluster. This pipeline automates the process of building, testing, and deploying applications. Each code commit triggers a series of automated tasks, including static code analysis, container image creation, and security checks. The pipeline ensures that only tested and verified code is promoted to production environments, significantly reducing deployment risks.

	\item \textbf{Kubernetes Clusters}: Two Kubernetes clusters form the core of the lab’s infrastructure. The primary cluster hosts the CI pipeline and containerized applications, while the secondary cluster is dedicated to machine learning experiments. These clusters are designed to be flexible and scalable, enabling rapid prototyping and testing of new workloads.

	\item \textbf{Security Tools and Policies}: Security is integral to the lab’s architecture. Tools such as \href{https://docs.bridgecrew.io/docs}{Bridgecrew} and \href{https://kyverno.io/}{Kyverno} enforce ``Security as Code'' and ``Policy as Code'' practices, ensuring that all configurations and deployments adhere to industry standards. Automated security scans are performed on code, configurations, and container images, reducing vulnerabilities across the development lifecycle.

	\item \textbf{Monitoring and Observability}: The lab integrates monitoring and observability tools to provide real-time insights into system performance and security. Metrics and logs are collected from Kubernetes clusters, CI pipelines, and containerized applications, allowing engineers to identify and resolve issues proactively.

	\item \textbf{Training and Experimentation}: As a training platform, the lab is designed to simulate real-world scenarios, providing engineers with hands-on experience in managing cloud-native infrastructures. The inclusion of advanced tools and workflows ensures that users can experiment with cutting-edge technologies while adhering to best practices.
\end{enumerate}

\justifying
The Cloudlab serves as a versatile platform that aligns theoretical research with practical applications. By integrating the components described above, the lab not only enhances operational resilience but also fosters innovation, making it a critical resource for advancing cloud-native security and scalability.

\ptitle{GitHub Repositories}

\justifying
The code base for the lab is broken down into several GitHub repositories, more or less around the functional area.

\begin{table}[ht]
	\centering
	\resizebox{\columnwidth}{!}{
		\begin{tabular}{| p{0.35\linewidth} | p{0.6\linewidth} |} \hline
			\cellcolor{myblue}\textcolor{white}{Repo Name} & \cellcolor{myblue}\textcolor{white}{Purpose}                                                                                     \\\hline
			\texttt{bot-cloudbot}                                   & A custom Python 3.9 GCP Cloud Function for GitHub pull request task automation.                                                  \\\hline
			\texttt{bot-ml-container}                               & An experimental containerized machine learning model.                                                                            \\\hline
			\texttt{gke-cluster}                                    & Infra as Code files for GKE cluster, YAML files for \href{https://tekton.dev/}{Tekton CI pipeline}. CN Series firewall nodes.    \\\hline
			\texttt{ps-containerizer}                               & An ``invisible shim'' with a Docker image for each ``public cloud'' VM-Series Terraform module development repo. Allows PRs to be
			ingested into Tekton CI pipeline without any integrations with the source repository.                                                                                             \\\hline
			\texttt{ps-devsecops-alpha}                             & IaC files for Alpha K8s cluster.                                                                                                 \\\hline
			\texttt{python-project-template}                        & Python project template for writing serverless code in AWS Lambda and GCP cloud functions.                                       \\\hline
		\end{tabular}
	}
	\caption{Project Codebase - GitHub Repositories}
	\label{mytable:1}
\end{table}

\vspace{2mm}
\ptitle{GitHub Actions}
\vspace{2mm}

\justifying
GitHub repositories that have been ``on-boarded'' to the project have certain ``actions'' included.

\begin{raggedright}
	\begin{enumerate}
		\item \href{https://docs.bridgecrew.io/docs}{Bridgecrew} is used to scan all commits to all open pull requests.
		\item Whitesource Renovate is used to track keep project dependencies up to date and secure.
		\item Another GitHub action defines the parameters of the GCP project and helps the ``ps-cloudbot'' with pull request  maintenance tasks.
	\end{enumerate}
\end{raggedright}
\vspace{2mm}

\justifying
Note that there is also a {webhook} to make the CI pipeline
aware of each commit and kick off a test and build cycle.
\vspace{2mm}
\justifying
\ptitle{Kubernetes Clusters}
\vspace{2mm}
\justifying
Two clusters are deployed with the Google Kubernetes Engine (GKE). The ``gke'' cluster hosts the \href{https://tekton.dev/}{Tekton CI pipeline}.
The ``alpha'' cluster is used for machine learning experimentation.

\justifying
Pipeline runs can be viewed and managed through a graphical interface that is well suited for development teams. There is
also the ability to manage pipeline runs and their requisite tasks using standard command line tooling.

\begin{mybox}{Example pipeline command}
franklin ~/gke: tkn pr ls  
NAME              STARTED        DURATION     STATUS
gh-pr-run-ncwrr   1 hour ago     1 minute     Succeeded
gh-pr-run-vhrvs   2 hours ago    1 minute     Succeeded
gh-pr-run-q6szq   3 hours ago    1 minute     Succeeded
gh-pr-run-8vn22   6 hours ago    1 minute     Succeeded
gh-pr-run-h7twc   14 hours ago   17 seconds   Failed
gh-pr-run-74wm7   21 hours ago   1 minute     Succeeded
gh-pr-run-tqlfg   1 day ago      1 minute     Succeeded
gh-pr-run-xs52f   1 day ago      1 minute     Succeeded
gh-pr-run-xtdz6   1 day ago      1 minute     Succeeded
gh-pr-run-w2zn7   1 day ago      1 minute     Succeeded

\end{mybox}
\vspace{2mm}

\ptitle{GCR Private Container Repository}
\vspace{2mm}

\justifying
To date there are about 15 GitHub repositories that are integrated with the CI pipeline outlined in this paper. Each commit
to a pull request causes the CI pipeline to generate a Docker image. These Docker images
are \href{https://cloud.google.com/container-registry/}{stored in a private GCR location}.
\vspace{2mm}

\ptitle{Panorama Management}
\vspace{2mm}

\justifying
Panorama is a key component in deployment and maintenance of Kubernetes clusters and CN Series firewalls. Currently there are
two virtual Panorama devices deployed in a high availability (HA) configuration.

\justifying
Palo Alto Networks has historically maintained serverless functions, for example in AWS Lambda, for firewall ``auto-scaling'' tasks. This has since been
replaced by a set of official Panorama ``plug-ins'' that can be downloaded to PanOS devices. Lambda and Cloud functions
are still frequently used to augment the capabilities of this new family of plug-ins.
\vspace{2mm}

\section{\label{sec:SRV}Serverless Functions}
\vspace{2mm}

\justifying
Google Cloud offers \href{https://cloud.google.com/functions/docs/concepts/overview}{Cloud Functions}, among other serverless computing offerings.
This is a good set of patterns to learn and comes up often
in security infrastructure work since organizations can use it to run jobs at a lower cost. Securing the
webhooks and connection between cloud functions and the VPC the GKE cluster is in is an important
factor in deployments. Making XML or other sorts of API calls is often used to pass data between management platforms, ticketing systems,
custom applications, and so on.
\vspace{2mm}

\subsection{Github Webhook}
\vspace{2mm}

\justifying
The Github webhook can be combined with a set of GCP credentials stored as a ``repository secret'' in
a repository. Available is a powerful and flexible
method to combine systems into a more intricate Continuous Integration and testing pipeline.

\begin{figure}[H]
\centering
	\includegraphics[width=0.8\linewidth]{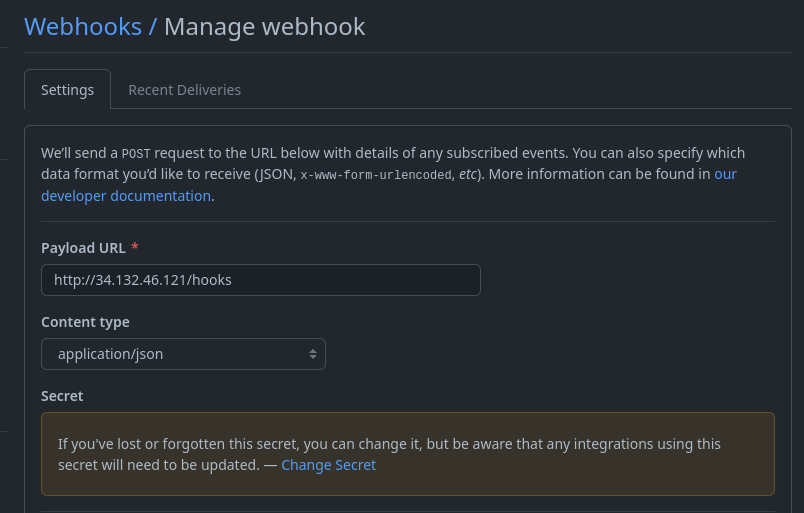}
	\caption{Webhook configuration settings in Github}
	\label{fig:pr3}
\end{figure}

\begin{figure}[H]
\centering
	\includegraphics[width=0.8\linewidth]{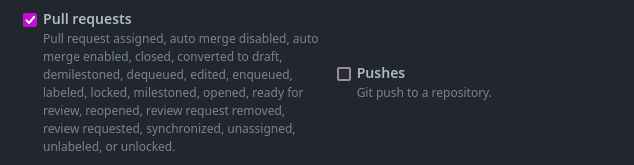}
	\caption{Check only this box in the webhook settings}
	\label{fig:pr4}
\end{figure}

\subsection{Connecting Serverless to GKE}
\vspace{2mm}

\justifying
A network peering is created between the Cloud Function and the VPC that the GKE cluster resides in.
This allows the serverless function application to make calls to a containerized deployment, such as
a Node application.
\vspace{2mm}

\begin{mybox}{Example YAML for a cloudbot service}

apiVersion: v1
kind: Service
metadata:
  name: cloudbot-service
  namespace: ci-build
  annotations:
    cloud.google.com/load-balancer-type: "Internal"
  labels:
    app: cloudbot-service
spec:
  type: LoadBalancer
  selector:
    app: cloudbot
  ports:
  - port: 80
    targetPort: 8089
    protocol: TCP
\end{mybox}

\begin{mybox}{Example deployment YAML for a cloudbot}

apiVersion: apps/v1
kind: Deployment
metadata:
  creationTimestamp: null
  labels:
    app: cloudbot
  name: cloudbot-deployment
  namespace: ci-build
spec:
  replicas: 3
  selector:
    matchLabels:
      app: cloudbot
  strategy: {}
  template:
    metadata:
      creationTimestamp: null
      labels:
        app: cloudbot
    spec:
      nodeSelector:
        env: build
      containers:
        - name: cloudbot
          image: gcr.io/gcp-gcs-pso/build-pod
          imagePullPolicy: Always
          volumeMounts:

            - name: nfs-volume-1
              mountPath: "/data"
      volumes:
        - name: nfs-volume-1
          persistentVolumeClaim:
            claimName: nfs-pvc
\end{mybox}

\begin{mybox}{Example Dockerfile for a cloudbot pod in GKE}

COPY ["app.js", "package.json", "package-lock.json", "./"]

RUN npm install --production

COPY . .

ENTRYPOINT ["node", "app.js"]
\end{mybox}
\vspace{2mm}

\section{\label{sec:CI}Continuous Integration}
\vspace{2mm}

\justifying
Ensuring the security of both internal and external build pipelines, as well as performing comprehensive scans on work products traversing these pipelines, is a critical aspect of modern cloud-native operations. These measures are essential to maintain the integrity, reliability, and security of the development and deployment processes. In alignment with these objectives, a fully operational Cloud Native Continuous Integration (CI) pipeline has been meticulously designed and implemented within the lab environment.

The implemented CI pipeline is designed to handle a wide range of repository types, whether they are public repositories or private ones that require secure authentication credentials. This flexibility ensures that organizations can leverage the pipeline for diverse projects, enabling seamless integration with existing workflows and infrastructure. The lab uses a specialized repository, referred to as the ``containerizer'' repository, to streamline the process of collecting files from source repositories, preparing them for deployment, and packaging them into secure and reliable container images.

\justifying
The ``containerizer'' repository acts as a critical intermediary, bundling source files with essential components such as command-line tools, test cases, and additional dependencies. These elements are necessary to validate and prepare the application for deployment. Once all components have been processed and verified, the pipeline generates a Docker image and securely stores it in the \texttt{gcr.io} container registry. This repository not only ensures a centralized location for managing container images but also includes features like automated vulnerability scanning, providing an added layer of security to the deployment lifecycle.

A pipeline run is the foundational unit of operation in this system. It represents a sequence of automated tasks, including testing, building, and packaging, that must be completed successfully to produce a deployable artifact. These tasks are orchestrated and executed using \href{https://tekton.dev/}{Tekton}, an open-source framework for creating cloud-native CI/CD pipelines. Tekton provides a scalable and flexible approach to managing these pipelines, allowing engineers to define tasks and workflows as code. This enables greater transparency, reproducibility, and adaptability to evolving project requirements.

\justifying
The CI pipeline offers several key capabilities and benefits:

\begin{enumerate}
	\item \textbf{Automated Testing and Validation}: Each code commit triggers a series of automated tests to validate functionality, security, and compliance with predefined standards. This ensures that only high-quality code progresses through the pipeline.
	
	\item \textbf{Dynamic Source Code Integration}: The pipeline seamlessly integrates with source repositories, whether public or private, enabling organizations to incorporate multiple contributors and projects without compromising security.

	\item \textbf{Comprehensive Security Scans}: The CI pipeline incorporates tools such as \href{https://docs.bridgecrew.io/docs}{Bridgecrew} to perform ``Security as Code'' scans, identifying and mitigating vulnerabilities in the codebase and container images.

	\item \textbf{Containerization and Image Management}: Leveraging the ``containerizer'' repository, the pipeline produces Docker images that are ready for deployment. These images include all necessary tools, libraries, and configurations, ensuring consistency across environments.

	\item \textbf{Storage and Accessibility}: All generated Docker images are securely stored in the \texttt{gcr.io} container registry. This centralized repository not only simplifies version management but also facilitates the secure distribution of images across environments.

	\item \textbf{Scalability and Modularity}: Tekton allows for the modular definition of tasks and workflows, enabling the CI pipeline to scale with the needs of the project. Engineers can add or modify tasks without disrupting the overall pipeline.

	\item \textbf{Enhanced Observability}: The pipeline includes logging and monitoring tools to provide real-time insights into its performance. This allows engineers to quickly identify and resolve issues during pipeline execution.

	\item \textbf{Seamless Deployment Readiness}: By the end of a successful pipeline run, a fully validated and packaged Docker image is produced, ensuring that deployments to production environments are secure, efficient, and reliable.
\end{enumerate}

\justifying
The role of \href{https://tekton.dev/}{Tekton} in orchestrating the CI pipeline cannot be overstated. Tekton’s task-based framework provides granular control over each step of the pipeline, allowing for extensive customization and optimization. Tasks within a pipeline can include activities such as linting, compiling, testing, security scanning, and packaging. Tekton also supports pipeline-as-code practices, enabling the storage of pipeline definitions in source repositories alongside application code. This ensures that the pipeline is version-controlled, auditable, and easily shareable.

Furthermore, Tekton’s integration with Kubernetes adds a layer of scalability and resilience to the pipeline. By running tasks as Kubernetes pods, the pipeline can dynamically allocate resources based on workload demands. This not only optimizes resource utilization but also ensures that pipeline tasks are isolated and secure.

In summary, the fully operational Cloud Native Continuous Integration pipeline implemented in the lab exemplifies best practices in modern DevOps. By combining automated testing, secure containerization, and dynamic orchestration through Tekton, the pipeline serves as a powerful tool for accelerating development while maintaining the highest standards of security and reliability. Its design and capabilities demonstrate the transformative potential of cloud-native technologies in streamlining CI/CD processes, making it an indispensable component of the lab's infrastructure.

\begin{figure}[H]
\centering
	\includegraphics[width=0.8\linewidth]{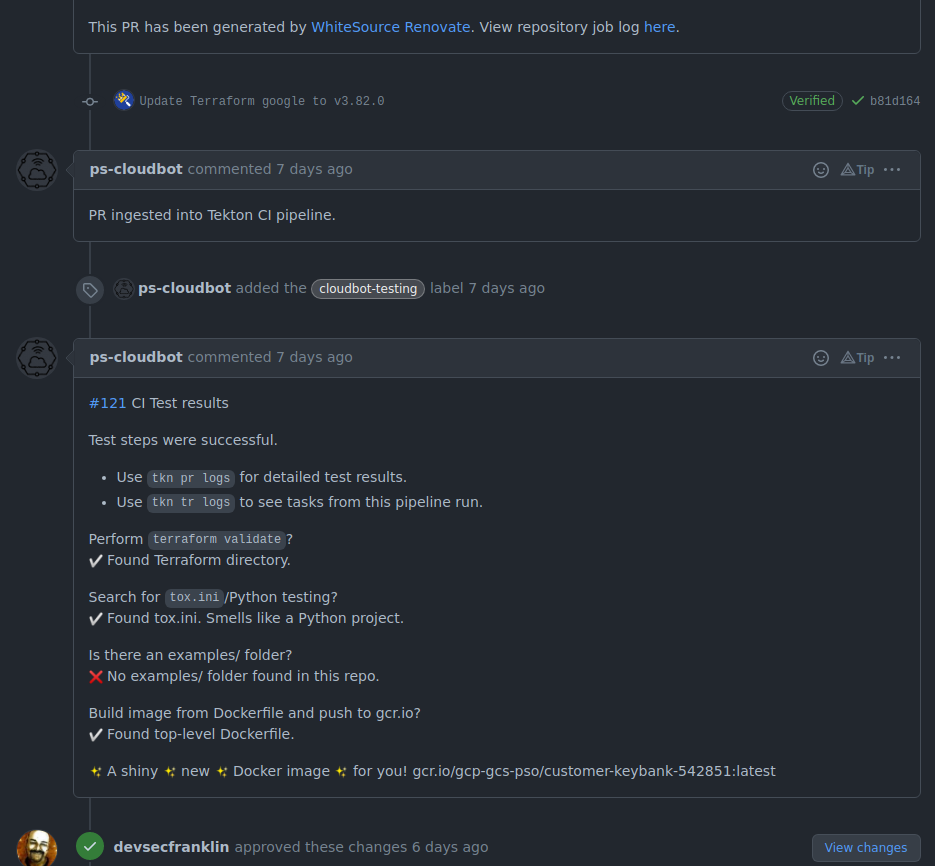}
	\caption{Tekton automated pipeline run results in a GitHub comment}
	\label{fig:pr1}
\end{figure}

Consider figure \ref{fig:pr1}. The ``Renovate'' bot detects an out-of-date or insecure dependency. A pull request is opened by the Renovate bot.
Next, the ``ps-cloudbot'' GCP cloud function is notified of the new pull request via webhook. The second bot places a label on the pull request to indicate
it is performing administrative actions on the pull request. The cloud function might perform other actions, such as assigning the pull request to a certain user,
adding certain users as reviewers, providing documentation, and so on. In this example, the cloud function adds a comment to the pull request to inform
project members that it has been accepted by the CI pipeline. A set of test cases based ona  certain technology or functional area is executed, and the
results are returned to the pull request in a second comment. There is also a link to the container image in the container registry. Although the pull
request is merged manually in this instance, the workflow could be modified to approve and merge the pull request with no human interaction whatsoever.

\begin{figure}[H]
\centering
	\includegraphics[width=0.8\linewidth]{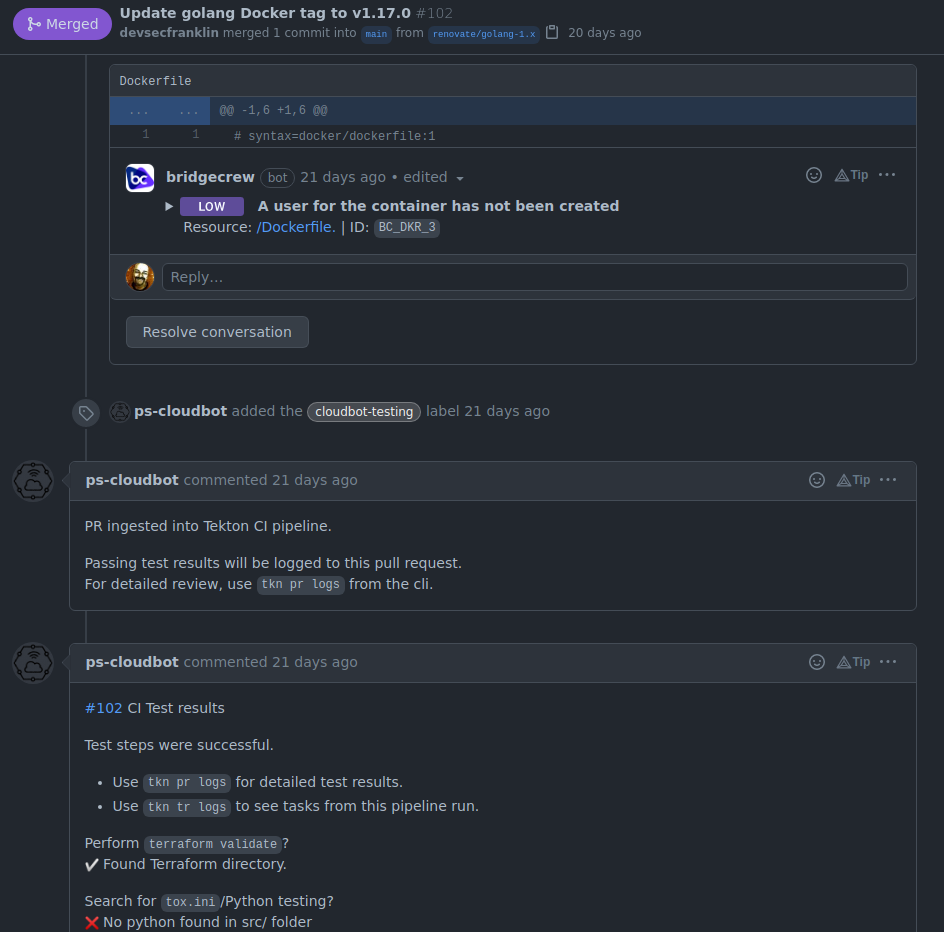}
	\caption{Bridgecrew integration with pull request automation}
	\label{fig:pr2}
\end{figure}

In addition to the ability to scan for dependencies that are out of date or have vulnerabilities of note, automated security scanning
of the code base can be performed as seen in figure \ref{fig:pr2}. As before, the Renovate bot has detected the use of an out-of-date Docker base image for Golang. Because
the Dockerfile is updated as part of this pull request, \href{https://docs.bridgecrew.io/docs}{Bridgecrew} scans the file and triggers misconfigurations that may lead to security issues.
The results of the scans and the issues found are noted in the pull request comments. Automated remediation and merging of these issues may be possible in
some cases.

\section{\label{sec:conclusion}Conclusion}

\justifying
The development and implementation of the Cloudlab project represent a significant step forward in advancing cloud-native security research and training. By combining state-of-the-art tools, methodologies, and infrastructure, the lab provides a robust platform for exploring modern IT challenges, fostering innovation, and equipping engineers with practical expertise. While the lab has already achieved substantial progress in areas such as container security, Policy as Code, and Continuous Integration pipelines, the potential for future enhancements remains vast.

Looking ahead, the lab aims to address several ambitious goals that will further expand its scope and capabilities. These goals include:

\begin{enumerate}
	\item \textbf{Orchestration of Multi-Cloud Infrastructure Builds and Deployments}: As organizations increasingly adopt multi-cloud strategies to optimize resource utilization and ensure redundancy, the lab seeks to explore advanced orchestration techniques. This includes leveraging tools like \href{https://crossplane.io/docs/v1.3/}{Crossplane}, which allows seamless integration and management of resources across multiple cloud providers. By achieving this, the lab can serve as a testbed for developing scalable, resilient, and interoperable multi-cloud solutions.

	\item \textbf{Expanding Knowledge of Infrastructure as Code (IaC) Languages}: While tools like Terraform (using HCL) have become synonymous with IaC, the lab aims to broaden its horizons by delving into alternative IaC languages and frameworks. For instance, \href{https://www.pulumi.com/}{Pulumi} offers a programming-language-based approach to IaC, enabling the use of familiar languages like Python, JavaScript, and Go for defining and managing infrastructure. Exploring Pulumi will allow the lab to compare and evaluate the strengths and use cases of different IaC paradigms.

	\item \textbf{Conducting Red Team Exercises}: Security remains at the heart of the Cloudlab project, and one of the key future goals involves ``red teaming'' the lab itself. By conducting dynamic application security testing (DAST) exercises, the lab can identify vulnerabilities, validate its defenses, and simulate real-world attack scenarios. These exercises will not only enhance the lab's security posture but also provide invaluable insights into how cloud-native systems can be hardened against evolving threats.
\end{enumerate}

\justifying
These future objectives highlight the lab’s commitment to continuous improvement and adaptability in an ever-changing technological landscape. By orchestrating multi-cloud environments, expanding expertise in IaC languages, and proactively addressing security vulnerabilities, the lab will remain at the forefront of cloud-native innovation.

\justifying
In conclusion, the Cloudlab project exemplifies the transformative potential of cloud-native technologies. It bridges the gap between theoretical knowledge and practical application, empowering researchers and engineers to tackle the complexities of modern IT ecosystems. With its robust infrastructure, innovative methodologies, and ambitious future goals, the lab not only supports groundbreaking research but also fosters a culture of learning and collaboration. As the lab evolves, it is poised to play a pivotal role in shaping the future of secure, scalable, and adaptive cloud-native environments.

\nocite{*}
\bibliographystyle{IEEEtran}
\bibliography{mybib.bib}

\end{document}